\documentclass[journal]{IEEEtran}
\usepackage{cite}

\ifCLASSINFOpdf
  \usepackage[pdftex]{graphicx}
  \DeclareGraphicsExtensions{.pdf,.jpg,.png}
\else
\fi
\usepackage{amsmath}
\usepackage{amssymb}

\usepackage{url}

\usepackage{siunitx}

\usepackage[draft,nomargin,inline,author=]{fixme}
\fxsetup{theme=color}
\definecolor{fxwarning}{rgb}{0.8,0.0000,0.0000}

\usepackage{booktabs}

\usepackage{adjustbox}

\newcommand*\around{{\raise.17ex\hbox{$\scriptstyle\mathtt{\sim}$}}}

\usepackage{multirow}

\begin{document}
%
\title{Noise Analysis of Photonic Modulator Neurons}
%
%
%

\author{Thomas~Ferreira~de~Lima\textsuperscript{*},
        Alexander~N.~Tait,
        Hooman~Saeidi,
        Mitchell~A.~Nahmias,
        Hsuan-Tung~Peng,
        Siamak~Abbaslou,
        Bhavin~J.~Shastri,
        and Paul~R.~Prucnal%
\thanks{\textsuperscript{*}Corresponding author: tlima@princeton.edu}
\thanks{T.F.L., A.N.T., H.S., M.A.N., H.-T.P., S.A., B.J.S. and P.R.P. are with the Department of Electrical Engineering, Princeton University, Princeton, NJ 08544, USA.}%
\thanks{B.J.S. is with the Department of Physics, Engineering Physics \& Astronomy, Queen's University, Kingston, ON KL7 3N6, Canada.}%
\thanks{A.N.T. is currently with the Physical Measurement Laboratory, National Institute of Standards and Technology, Boulder, CO 80305, USA.}}
\maketitle

{\color{black}
\begin{abstract}
Neuromorphic photonics relies on efficiently emulating analog neural networks at high speeds. Prior work showed that transducing signals from the optical to the electrical domain and back with transimpedance gain was an efficient approach to implementing analog photonic neurons and scalable networks. Here, we examine modulator-based photonic neuron circuits with passive and active transimpedance gains, with special attention to the sources of noise propagation. We find that a modulator nonlinear transfer function can suppress noise, which is necessary to avoid noise propagation in hardware neural networks. In addition, while efficient modulators can reduce power for an individual neuron, signal-to-noise ratios must be traded off with power consumption at a system level. Active transimpedance amplifiers may help relax this tradeoff for conventional p-n junction silicon photonic modulators, but a passive transimpedance circuit is sufficient when very efficient modulators (i.e. low C and low V-pi) are employed.
\end{abstract}
}


%
\IEEEpeerreviewmaketitle

\section{Introduction}

\IEEEPARstart{T}{he gap} between current computing capabilities and current computing needs ushered research in the field of neuromorphic computing~\cite{Merolla:2014,Furber:14,Benjamin:14,Meier:15,Davies:18}. This new field aims to bridge the gap between the energy efficiency of von Neumann computers and the human brain~\cite{Marr:13,Hasler2013}. As a consequence, this thrust spawned research into novel brain-inspired algorithms and applications uniquely suited to neuromorphic processors. These algorithms attempt to solve artificial intelligence tasks in real-time while using less energy.
We posit that we can make use of the high parallelism and speed of photonics to bring the same neuromorphic algorithms to applications requiring multiple channels of multi-gigahertz analog signals, which digital processing struggles to process in real-time.

By combining the high bandwidth and parallelism of photonic devices with the adaptability and complexity attained by methods similar to those seen in the brain, photonic processors have the potential to be at least ten thousand times faster than state-of-the-art electronic processors while consuming less energy per computation~\cite{FerreiradeLima2016}. An example of such an application is nonlinear feedback control; a very challenging task that involves computing the solution of a constrained quadratic optimization problem in real time. Neuromorphic photonics can enable new applications because there is no general-purpose hardware capable of dealing with microsecond environmental variations~\cite{FerreiradeLima2019}.

These benefits can be accomplished by use of wavelength-division multiplexing (WDM), which explores the enormous bandwidth of optical waveguides (\around THz).
Integrated neuromorphic circuits based on WDM can now be manufactured using silicon photonics platforms.
Tait et al. recently demonstrated a way of performing neural computations on WDM signals via a photodetector directly driving a modulator~\cite{Tait:18:mrrMod}, which is capable of integrating hundreds of wavelength channels carrying gigahertz signals.
We call this an O/E/O-based photonic neuron (Fig.~\ref{fig:oeo diagram}).

In this architecture, photonic neurons output optical signals with unique wavelengths. These are multiplexed into a single waveguide and broadcast to all others, weighted, and photodetected. Each connection between a pair of neurons is configured independently by one MRR weight~\cite{Tait:16scale,Tait:16multi}, and the WDM carriers do not mutually interfere when detected by a single photodetector.
Consequently, the physics governing the neural computation is fully analog and does not require any logic operation or sampling, which would involve serialization and sampling. Thus they exhibit distinct, favorable trends in terms of energy dissipation, latency, cross-talk and bandwidth when compared to electronic neuromorphic circuits~\cite[Sec. 5]{FerreiradeLima2016}.

But the same physics also introduce new challenges, especially reconfigurability, integration, and scalability.
{\color{black}
Information carried by photons is harder to manipulate compared to electronic signals, especially nonlinear operations and memory storage. Photonic neurons described here solve that problem by using optoelectronic components (O/E/O), which can be mated with standard electronics providing reconfigurability. However, neuromorphic photonic circuits are challenging to scale up because they do not benefit from digital information, memory units and a serial processor, and therefore requires a physical unit for each element in a neural network, increasing size, area and power consumption. Here, integration costs must also be considered, since the advantages of using analog photonics (high parallelism and high bandwidth) must outweigh the costs of interfacing it with digital electronics (requiring both O/E and A/D conversion). The optimal cost-benefit tradeoff can be computed for particular applications by engineers, but one factor that has not been addressed in prior literature is the accumulation of noise in O/E/O neuromorphic analog links.
}
This paper will provide a quantitative study in this dimension of O/E/O neural networks.

\subsection{Analog Processing}

The main advantage of this O/E/O approach is that it solves the problem of transferring energy between WDM signals into a single wavelength output. This property enables O/E/O neurons to be networked together via broadcast waveguides.

\begin{figure}[ht]
\centering
\includegraphics[width=\linewidth]{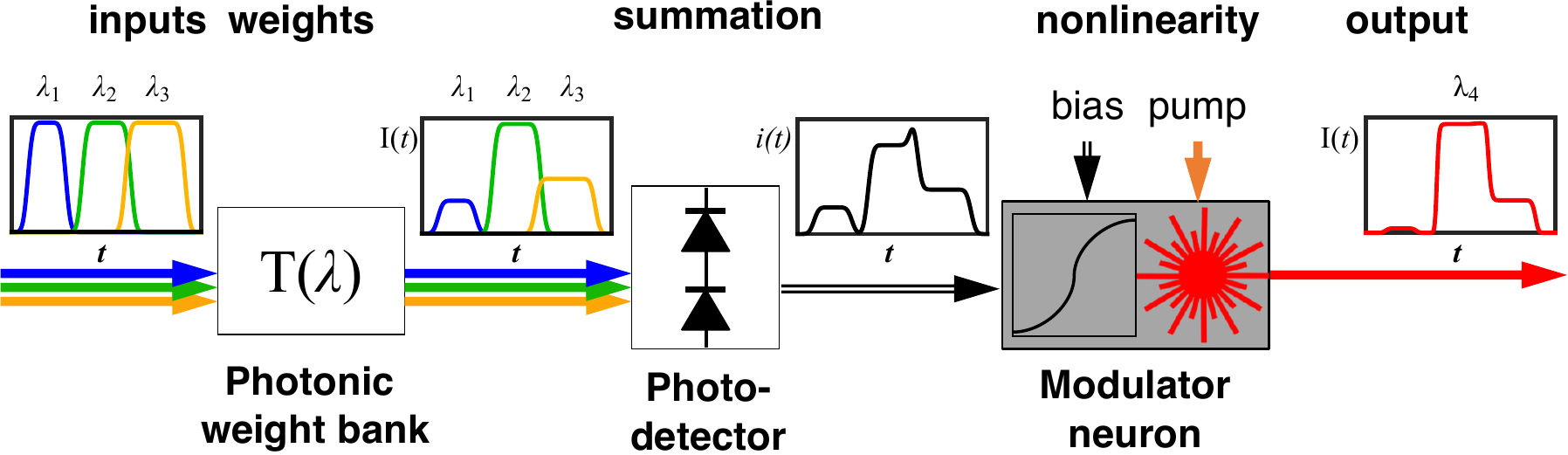}
\caption{Diagram of an O/E/O photonic neuron.}
\label{fig:oeo diagram}
\end{figure}

The fact that lightwaves are transduced into electrical currents and back is very advantageous for implementing nonlinear operations, compared to all-optical strategies. If all devices worked perfectly, this scheme would work even with extremely low power and voltage levels. At the low-energy limit, one photon transforms into a electron-hole pair, whose charge can be used to modulate the transmission of an electro-optic device, sourcing a number of photons at the output of the circuit (Fig.~\ref{fig:oeo diagram}).
This mode of operation would require lossless optics, coupled with very sensitive photodetectors and modulators.

Because of such loss and inefficiency, we need to operate the neuron with stronger signals. In addition, due to the analog nature of this communication scheme, it is not immune to noise accumulation. Ultimately it is the shot noise and the noise of circuit components that {\color{black} currently} prevents driving these O/E/O photonic neurons with quantum-scale power levels {\color{black} at room temperature}.

\begin{figure}[ht]
\centering
\includegraphics[width=\linewidth]{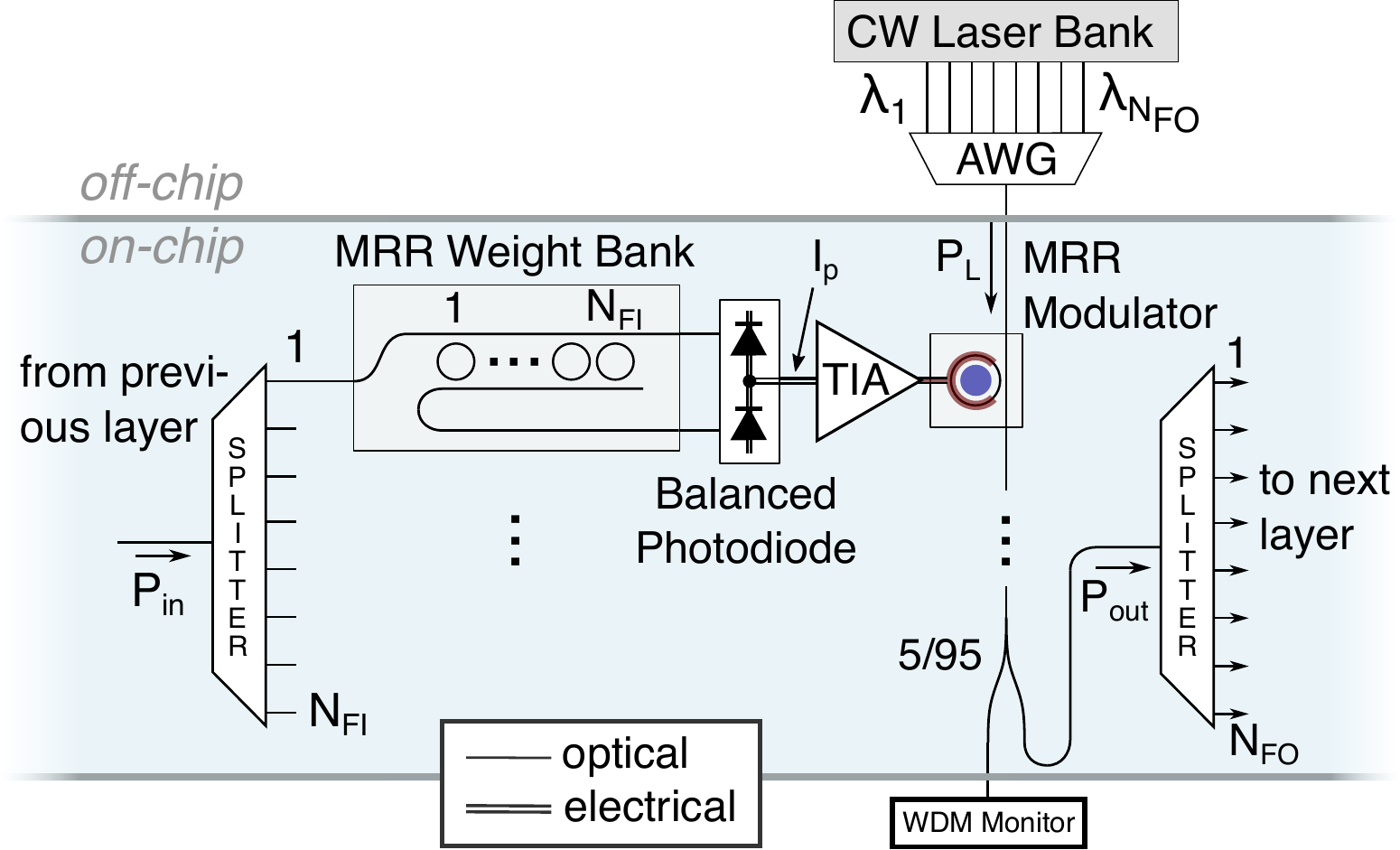}
\caption{Photonic neurons that receive inputs from one WDM broadcast medium and transmit into another, corresponding to the next layer. Distinct layers can thus reuse the same optical spectrum for broadcasting, much like a cellular telephone network reuses spectrum geographically. This strategy of using more waveguides for more layers can be extended, in principle, to an arbitrary depth.}
\label{fig:pnn schematic}
\end{figure}

In order for one layer to physically drive the next layer, enough laser power $P_L$ must be provided to compensate for loss and power splitting due to fan-out to the next layer ($N_\text{FO}$). Alternatively, electrical gain via a high transimpedance ($R_\mathrm{TIA}$) can provide the necessary amplification.
This ``gain cascadability'' condition can be written with the following inequality:
\begin{equation}
\label{eq:gain cascadability}
\underbrace{R_{\text{TIA}}}_{\text{electrical}}
\cdot
\underbrace{P_{L}}_{\text{optical}}
\cdot
\underbrace{\left(\frac{\mathrm{MD}}{V_{\text{pp}}}\right)}_{\text{mod. sens.}}
>
\frac{N_{\text{FO}}}{2T_{1/2}R_d\eta_{\text{pp}}}
\end{equation}
where $\mathrm{MD}$ is the modulation depth; $T_{1/2}$, mean transmission, $R_d$, photodiode's responsivity, and $\eta_{\text{pp}}$ is the optical point-to-point efficiency between connected neurons, {\color{black} representing excess optical loss between the output of one neuron and the input of the next, i.e. propagation loss and insertion loss of weighting devices and couplers}.

Equation~\ref{eq:gain cascadability} describes the relationship between different kinds of gain: electrical, optical, modulator sensitivity. We draw attention to the $R_{\text{TIA}}/V_{\text{pp}}$ ratio (transimpedance over peak-to-peak voltage), which represents the \emph{neuron's sensitivity}, {\color{black} because it quantifies how much photocurrent is necessary to effect a full amplitude swing in the modulator.} Since optical pump power ($P_L$) is an expensive resource, it is desirable to maximize this sensitivity in order to minimize overall power consumption. However, higher sensitivity comes at a price, as noise accumulation degrades the signal-to-noise ratio at the output (Sec.~\ref{sec:1 noise-propagation}).

\subsection{Suppressing Noise Accumulation}
Without careful design, analog circuits with long chains of cascaded neurons can accumulate and amplify noise, eventually burying signals under the noise floor.
The optoelectronic devices shown in Fig.~\ref{fig:pnn schematic} are mostly linear and noisy, resulting in a signal-to-noise ratio (SNR) degradation, i.e. a noise factor greater than unity ($F>1$). Fundamentally, this happens because not only they linearly amplify noise as well as signals, they generate noise on top of the output. To counter that effect, we need a device that can amplify signals while decreasing noise. This can be achieved with a nonlinear device -- in our case, a modulator with a nonlinear transfer curve. The more nonlinear the modulator is, the more it can compensate for accumulating noise (Fig.~\ref{fig:mod transfer function}).

\subsection{Organization of the Paper}
In this paper, we will numerically analyze the noise as it propagates across a neural network composed of O/E/O neurons~(Sec.~\ref{sec:1 noise-propagation}). We will quantify the nonlinearity requirements for a neural circuit to suppress noise accumulation. We propose a simple experiment involving the cascadability property of the neuron which verifies that noise is properly suppressed.

Following the noise analysis, we show two options of constructing the O/E/O circuit: one with a passive electronic link, and the other with an active transimpedance amplifier which provides energy gain in the electronic domain~(Sec.~\ref{sec:2 technology-comparison}). The electronic gain can enhance the neuron's response for weak input signals.
The effect of this enhancement is a reduction of overall optical power levels in the circuit, leading to more energy efficiency.

\section{Noise Propagation in O/E/O Neurons}
\label{sec:1 noise-propagation}

Neural networks are known to be robust to noise~\cite{Lee1994}.
In fact, noise can be exploited to train neural networks when other optimization algorithms might fail~\cite{Abdelaziz2015,Bengio2013}. Noise originating in hardware was used to implement on-line learning to a VLSI neural network~\cite{Jayakumar1992}.


There are two methods for avoiding noise propagation across a network. The first involves a collective approach, using redundant neurons encoding {\color{black} correlated} information.
{\color{black} This is called population coding in neuroscience, and is it required by physiological neural networks to overcome the noise generated by individual neurons~\cite{Snippe1992,Seung1993}. In essence, the estimation error for information carried by $N$ neurons scales with $1/\sqrt{N}$~\cite{Seung1993}.
This concept has been adapted to machine learning and has been proven to mitigate noise in multilayer perceptron networks~\cite{Lee1994}, the category into which photonic neural networks described in this paper falls.
}
The second method, the focus of this paper, relies on every individual neuron to have noise suppressing circuit. This section studies the noise accumulation mechanisms within a single neuron and describes how a modulator's nonlinearity can suppress noise accumulation.

\subsection{Modulator Nonlinearity as Noise Trimmer}

\begin{figure}[ht]
\centering
\includegraphics[width=.7\linewidth]{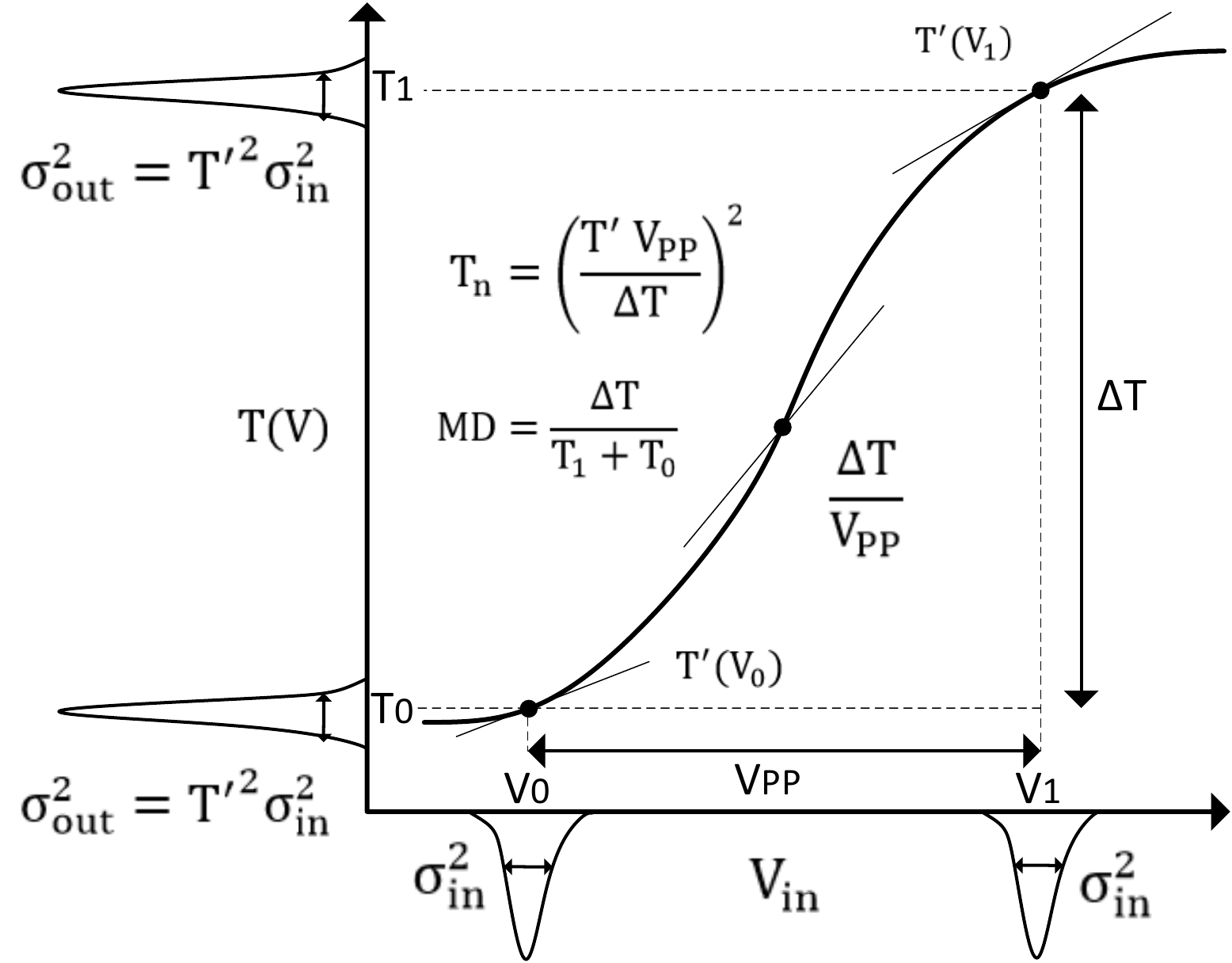}
{%
\fbox{\includegraphics[width=.22\linewidth]{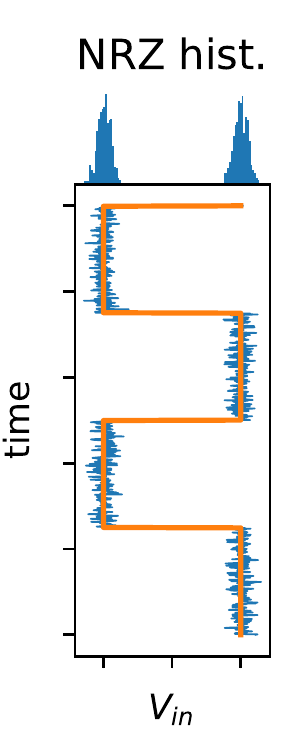}}%
}%
\caption{Modulator's transfer function showing the noise trimming principle. $V_{\text{pp}}$ and $T(V)$ represent the peak-to-peak voltage and transfer function of this modulator, respectively. $T(V)$ is assumed to be a symmetric S-shape, so $T'(V_0) = T'(V_1)$. Here, an NRZ signal with probability distribution shown in the x-axis can be transduced into an optical signal with lower noise (y-axis).
{\color{black} The inset illustrates how to extract the distribution empirically.}
}
\label{fig:mod transfer function}
\end{figure}

Consider non-return-to-zero (NRZ) modulated signals at the input and output of each neuron. Assume that the neuron is biased so that zeros and ones fall on each side of its S-shaped transfer function (Fig.~\ref{fig:mod transfer function}). Because the derivative of this function is relatively small in these regions, noise variance is reduced and the output looks ``cleaner'' than the input. This operating principle can be generalized to other modulation schemes and other transfer functions, but exploring all theoretical possibilities is beyond the scope of this paper.

\subsection{Sources of Noise in the OEO Link}
\begin{figure}[ht]
\centering
\includegraphics[width=.9\linewidth]{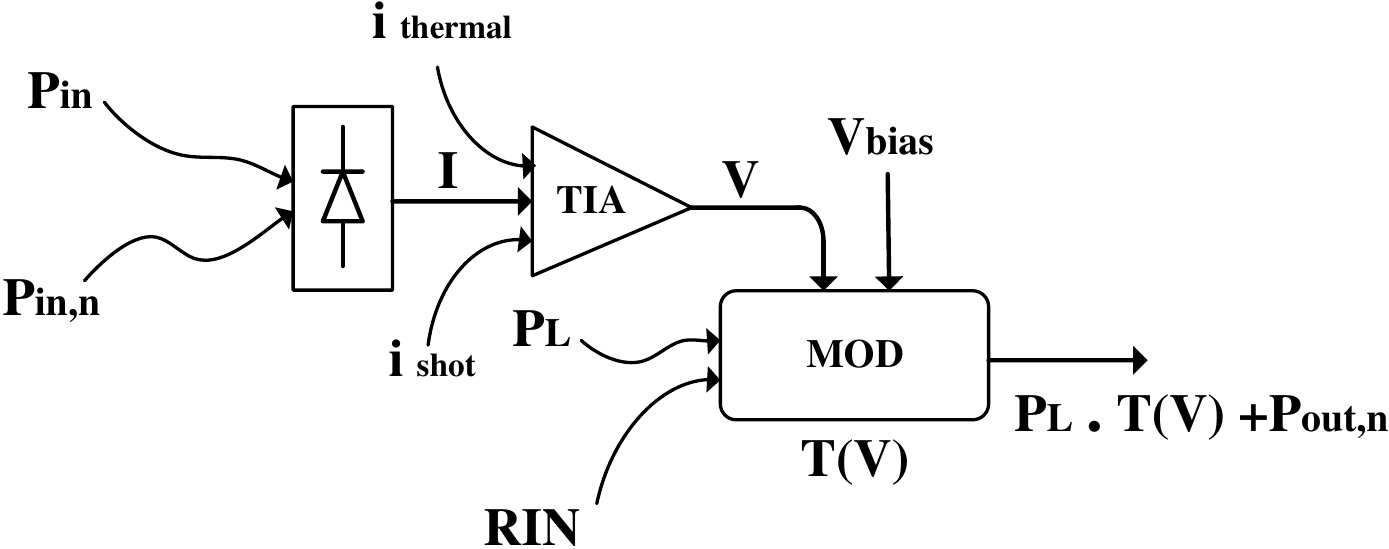}
\caption{Simplified neuron circuit showing the sources of noise in each step of the photonic link.}
\label{fig:noise sources}
\end{figure}

The accumulation of thermal noise, shot noise, amplifier noise, and relative intensity noise (RIN) must be counteracted by the modulator nonlinearity in order to guarantee that the SNR at the output equals the SNR at the input. This condition leads to the following equation:

\begin{align}
\label{eq:SNR}
\frac{1-T_n^2}{\mathrm{SNR}} =&
\frac{4T_n^2}{V_{\text{pp}}^2}
\biggl(
    \underbrace{\frac{4k_BT\Delta f R_{\text{TIA}}^2}{R_L}}_{\text{thermal noise}}
    +
    \underbrace{2q\Delta f V_{\text{pp}} R_{\text{TIA}}}_{\text{shot noise}}
    +
    \IEEEnonumber\\
    &+
    \underbrace{R_{\text{TIA}}^2 \Delta f I_{\text{TIA,n}}^2}_{V_{\text{TIA,n}}^2=\text{TIA noise}}
\biggr)
+
\underbrace{\mathrm{RIN}^2\left(1+\frac{1}{\mathrm{MD}^2}\right)}_{\text{RIN}}
\end{align}
where four sources of voltage noise (depicted in Fig.~\ref{fig:noise sources}) are balanced against a potentially very low term $T_n$.
The expression derivation, assumptions, and precise meaning of each term are described in Appendix~\ref{appendix:derivation}.

Equation~\ref{eq:SNR} shows that the SNR converges to a finite level which cannot be arbitrarily large, since the term on the right is always positive. That is expected since noise is generated at every stage. To increase the neuron's SNR, we need to decrease the terms on the right side of the equation, in particular $T_n^2$ (noise transmission) and $(R_{\text{TIA}}/V_{\text{pp}})$ (sensitivity).
The modulator's nonlinearity is very important in guaranteeing a high SNR. A completely linear modulator (leading to a noise transmission factor $T_n\rightarrow1$) results in a completely noisy signal ($\mathrm{SNR}\rightarrow 0$). This happens because noise increases more strongly than the signal at every stage, eventually reducing signal integrity at the infinite cascadability limit.
However, a completely nonlinear modulator ($T_n\rightarrow0$) results in a high quality signal ($\mathrm{SNR}\rightarrow (2\mathrm{RIN})^{-2}$), which can be over \SI{100}{dB} {\color{black} (assuming $\mathrm{RIN}=\SI{-160}{dB\per\Hz}$~\cite{Resneau2006})}.

\subsection{Noise vs. Gain Tradeoff}

Modulators in reality have intrinsic nonlinearities, but are often operated in their linear region, often limiting their extinction ratio. However, nonlinearities are not only necessary by the mathematics of the neural network but are also here exploited to suppress noise.
The microring resonator modulator, in particular, is known to have a Lorentzian-shape transfer function, which provides an ideal high-sensitivity S-shape for this application. Another candidate is a Mach-Zehnder modulator, with a sinusoidal transfer function. In either case, we expect their noise multipliers to lie strictly between 0 and 1.

Assuming $0 < T_n < 1$ and a fixed $V_{\text{pp}}$, then SNR can be increased by reducing transimpedance $R_{\text{TIA}}$ as much as possible, at the expense of greater power consumption (Eq.~\ref{eq:gain cascadability}). In other words, a higher quality signal requires a lower gain circuit, resulting in a higher power consumption. This tradeoff can be exploited to save energy in neural network applications for which SNR is not a critical factor. Section~\ref{sec:2 technology-comparison} shows a few realistic estimations for silicon photonic modulator neuron implementations.

\subsection{Autapse Test as Cascadability Standard}

\begin{figure*}[ht]
\centering
\begin{adjustbox}{totalheight=120pt,valign=M}
\includegraphics[width=\linewidth]{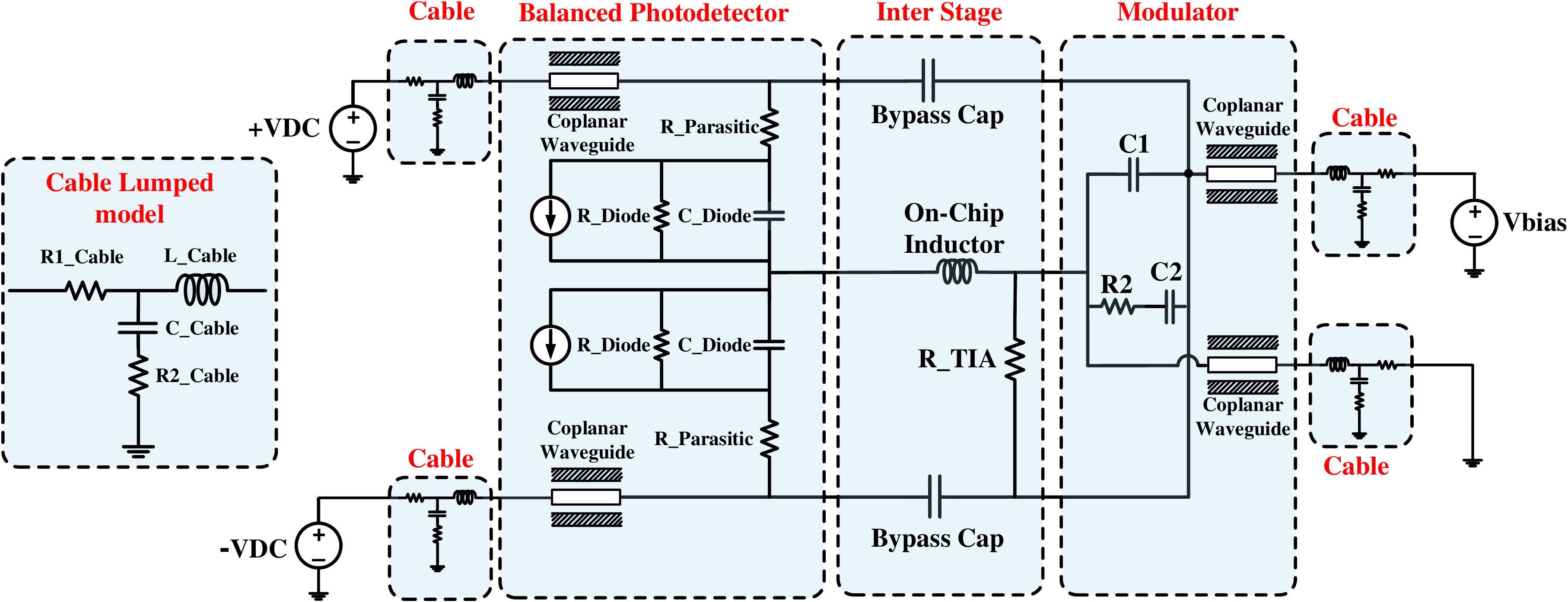}%
\end{adjustbox}
\begin{adjustbox}{totalheight=100pt,valign=M}
    \begin{tabular}{cc}%
    \toprule
    \multicolumn{2}{c}{\textbf{Cable Lumped Model}}\\\midrule
    $R_{1,2}$ & \SI{100}{\ohm}\\
    $C$ & \SI{22}{\pF}\\
    $L$ & \SI{220}{\nano\henry}\\
    \midrule
    \multicolumn{2}{c}{\textbf{Balanced PD Model}}\\\midrule
    $R_{\text{diode}}$ & \SI{17}{\kilo\ohm}\\
    $C_{\text{diode}}$ & \SI{20}{\fF}\\
    $R_{\text{parasitic}}$ & \SI{57}{\ohm}\\
    \bottomrule
    \end{tabular}%
\end{adjustbox}
\begin{adjustbox}{totalheight=100pt,valign=M}
    \begin{tabular}{cc}%
    \toprule
    \multicolumn{2}{c}{\textbf{Modulator Model}}\\\midrule
    $C_{\text{1}}$ & \SI{34.7}{\fF}\\
    $C_{\text{2}}$ & \SI{14.7}{\fF}\\
    $R_{\text{2}}$ & \SI{19.3}{\kilo\ohm}\\
    \midrule
    \multicolumn{2}{c}{\textbf{Inter Stage (TIA)}}\\\midrule
    $R_{\text{TIA}}$ & \SI{200}{\ohm}\\
    $C_{\text{bypass}}$ & \SI{1}{\nano\F}\\
    $L_{\text{peak}}$ & \SI{5.1}{\nano\henry}\\
    \bottomrule
    \end{tabular}%
\end{adjustbox}
\caption{Circuit schematic of a silicon photonic neuron with a passive transimpedance circuit whose bandwidth is enhanced via inductive gain peaking~\cite{Novack2013}. The circuit parameters are typical of recent literature and have been experimentally verified~\cite{Lee2014,Xuan2014,Novack2013}.
\vspace{-15pt}
}
\label{fig:passiveTIA}
\end{figure*}

We use the notion of cascadability to verify whether a particular photonic neuron design can be scaled up to form large networks.
There are three kinds of cascadability: physical, gain and noise.
\paragraph{Physical} A photonic neuron device is physically cascadable if the nature of its output can be directly connected to another's input. For example, the neuron introduced in Fig.~\ref{fig:oeo diagram} outputs a lightwave of a single wavelength, while receiving a number of inputs in parallel at different wavelengths. The signals are carried in the amplitude envelope (not phase, or polarization) of both input and output lightwaves. Because of that, this neuron can be interconnected and form arbitrarily-large neural networks so long as the fan-in to each neuron is equal or less than the number of WDM channels it is designed to support.

\paragraph{Gain} Beyond being physically cascadable, each neuron must be able to provide enough {\color{black} optical} power to excite the next layer of neurons, if needed. Equation~\ref{eq:gain cascadability} provides an estimate of the gain cascadability condition for the worst case scenario: one neuron, alone, delivering enough optical energy to $N_{\text{FO}}$ other neurons. In this subnetwork, $N_{\text{FO}}$ neurons can ``replicate'' the output of the initial neuron, which allow for $N_{\text{FO}}$ subnetworks to process the input in parallel. In practical deep neural networks, however, multiple neurons share the burden of providing enough optical energy for the next layer. This metric can only be quantified if the shape and weight configuration of the neural network is known in advance, i.e. in the presence of an application benchmark, which is out of the scope of this article.

\paragraph{Noise} The other potential scalability limitation is noise accumulation. This is particularly important for deep networks. In the worst case scenario, the information contained in a signal fed to the first layer's input must survive uncorrupted as it goes through the remaining layers of the network, even in the presence of noise. The calculations leading to Eq.~\ref{eq:SNR} show that at the limit of infinitely deep neural networks, the SNR stabilizes to a certain value (solution of the equation) by balancing noise generation by the electronic {\color{black} O/E/O} link and the noise suppression by the modulator's nonlinearity.

\paragraph{Autapse Test}
A simple experiment can be constructed to demonstrate and quantify all three conditions: a self-connection, also referred to as an \emph{autapse}. With an autaptic connection with unity weight, the neuron emulates an infinite chain of neurons, where each connection delay $\tau$ represents a \emph{virtual} neuron. The autapse experiment thus allows to study infinite cascadability without producing an infinite chain of cascaded neurons.
In this experiment, an initial pulse perturbation is sent to the neuron at $t=0$, triggering an output pulse in response. This output pulse travels through the autapse and, provided the gain cascadability condition is met, excites another perturbation at $t=\tau$. The evolution of the pulse amplitude and shape at times $t=n\tau$ will determine whether this neuron has met both gain and noise cascadability conditions.
{\color{black}
This experiment emulates an infinite series of neurons connected on a one-to-one basis. However, it can also emulate a one-to-N connectivity pattern if the autapse weight is set to $1/N$, which represents a $1/N$ loss in optical power between consecutive layers. Many-to-one and many-to-many connectivity can be extrapolated from this test but not directly emulated. We also note that this experiment tests for indefinite cascadability, which might not be required in small neural networks.
}

Autapse experiments as described here were conducted in both modulator-based~\cite[Sec.~E]{Tait:18:mrrMod} and laser-based~\cite{Peng:19:CLEO} photonic neurons, but they focused on demonstrating physical and gain cascadability. Observing noise accumulation in an autapse is the next logical step in testing these devices.

\section{Transimpedance Amplifier in Silicon Photonics}
\label{sec:2 technology-comparison}

Noise suppression relies on using the modulator's intrinsic nonlinear transfer curve. But the nonlinearity is only observed if the voltage swing to the modulator is large enough (Fig.~\ref{fig:mod transfer function}). The suite of standard silicon photonic components today are based on Ge photodetectors and p-n junction index modulators which possess a capacitance on the order of tens of femtofarads and require a voltage swing of a few volts. This voltage swing is provided by a transimpedance amplifier (TIA), which transduces photocurrent into voltage swing. Equation~\ref{eq:gain cascadability} already showed us that increasing the transimpedance reduces the optical pump power for the neuron. But increasing it too much limits the bandwidth of the circuit (inversely proportional to $RC$). In Sec.~\ref{subsec:TIA:passive} we explore how the modulator's parameters affect this power-bandwidth tradeoff. Sec.~\ref{subsec:TIA:active} discusses how active TIA can be used to mitigate some of that tradeoff. We show that an active TIA is no longer necessary to maintain \SI{10}{\GHz} bandwidth for sub-femtofarad nanophotonic devices.

\subsection{Passive Transimpedance}
\label{subsec:TIA:passive}

An easy way to control the transimpedance in passive silicon photonic chips is to use a simple resistor in parallel with the modulator, whose value determines the transimpedance gain. This design is simple and works well, but the photodetector and modulator's junction capacitance add in parallel with the transimpedance value.
{\color{black}
The achievable bandwidth is determined by the dominant pole of the circuit ($\Delta f = 1 / 2\pi R_{\text{TIA}} C$) (see~Fig.~\ref{fig:passiveTIA}).
}
As a result, this limits how large the transimpedance can be, since the capacitances add up to \SI{50}{fF}, which for a \SI{10}{GHz} bandwidth corresponds to a maximum of $R_{\text{TIA}}=\SI{320}{\ohm}$. This is significant because large networks will require on the order of 100 parallel wavelengths in a single waveguide. Assuming a maximum safe power of \SI{100}{mW} per waveguide (avoiding nonlinear effects~\cite{Koos2007}), that gives us a maximum of \SI{1}{\mW} per wavelength, generating only \around\SI{0.3}{V} of swing at the modulator, far from the typical $V_\pi$\around\SI{2}{V} required in silicon modulators~\cite{Dong2009}.
Fig.~\ref{fig:passiveTIA} shows an implementation compatible with standard silicon photonic foundry chips.

In this case, gain cascadability can only be achieved if (a) one uses a modulator with $V_\pi$\around\SI{0.3}{V} or (b) with a smaller capacitance, or (c) one uses an active TIA component which can decouple the transimpedance from the modulator's capacitance, allowing for a higher gain with the same bandwidth.


The solution involving improving modulators is promising, as we are far from the fundamental limits of photonics~\cite{Sorger2018}. Efficient modulators are key to cope with increasing demand in data communications, so research in this direction abounds. Exotic materials such as graphene are being used to reduce the switching energy of nanophotonic modulators toward sub-fJ~\cite{amin2018attojoule}. Photonic crystals also offer an avenue for ultracompact O/E/O conversion. For example, Nozaki et~al.\cite{Nozaki:18:OEO} have demonstrated a nanophotonic (InP-based) O/E/O link with \SI{1.6}{fF} capacitance and \SI{25}{\kilo\ohm} transimpedance, with a voltage swing of \SI{0.5}{V}. \SI{40}{\uW} was sufficient to operate this O/E/O device.


\subsection{Active Transimpedance Amplifier}
\label{subsec:TIA:active}

\begin{figure}[ht]
\centering
\includegraphics[width=.9\linewidth]{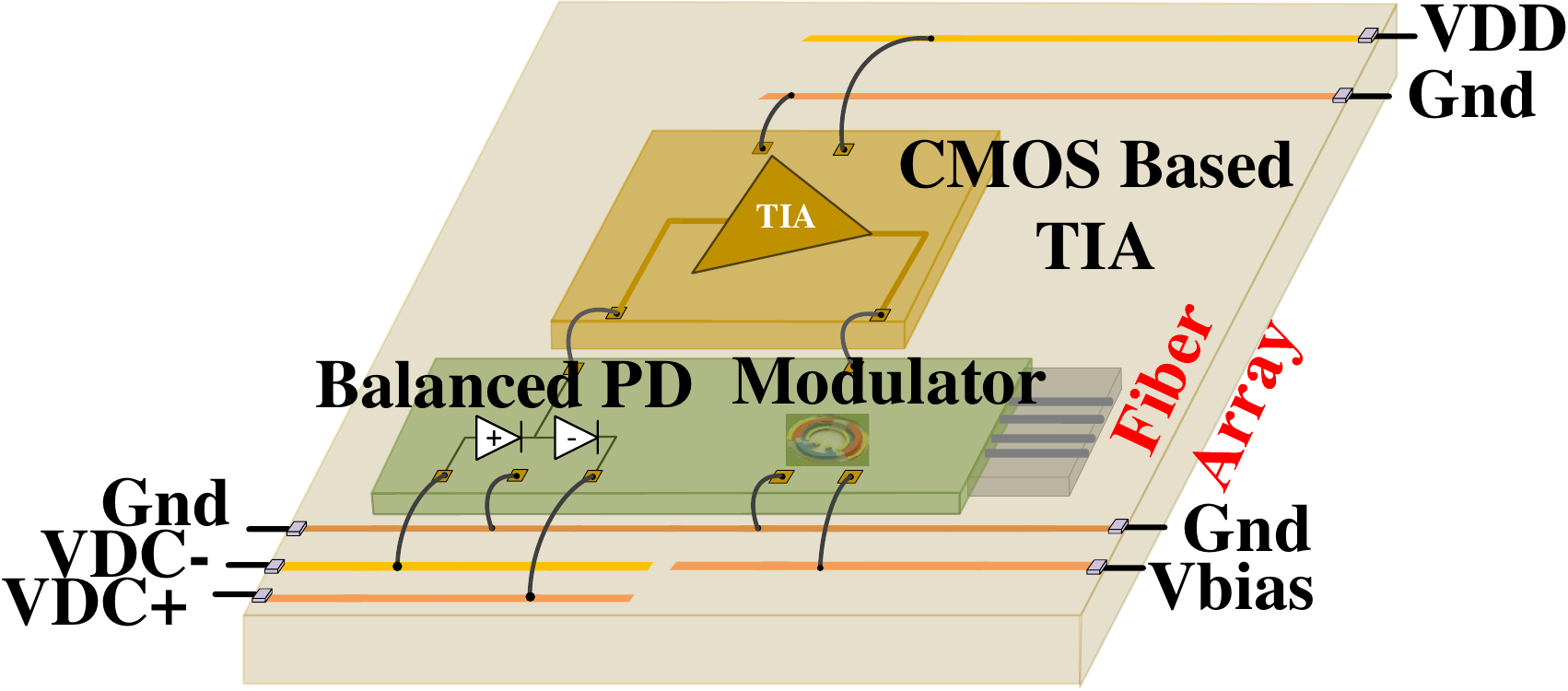}
\caption{Concept diagram of a silicon photonic integrated circuit packaged with a CMOS-based TIA. A flip-chip bonded alternative would yield similar electrical performance for {\color{black} the bandwidth of interest ($10\mathrm{GHz}$)}.}
\label{fig:activeTIA}
\end{figure}

Another way to provide gain without compromising bandwidth is to use an active TIA circuit~\cite{Kim2012,Jun-DeJin2008} instead of a RLC circuit. The TIA is designed to enhance the voltage swing of the modulator when the photocurrent is limited. In the short term, photonic integrated circuits can be coupled with CMOS-based TIAs via wirebonds or flip-chip bonding (Fig.~\ref{fig:activeTIA}). In the long run, however, these may be homogeneously integrated on the same chip, via a zero-change platforms~\cite{Stojanovic2018}.

\begin{figure}[ht]
\centering
\includegraphics[width=\linewidth]{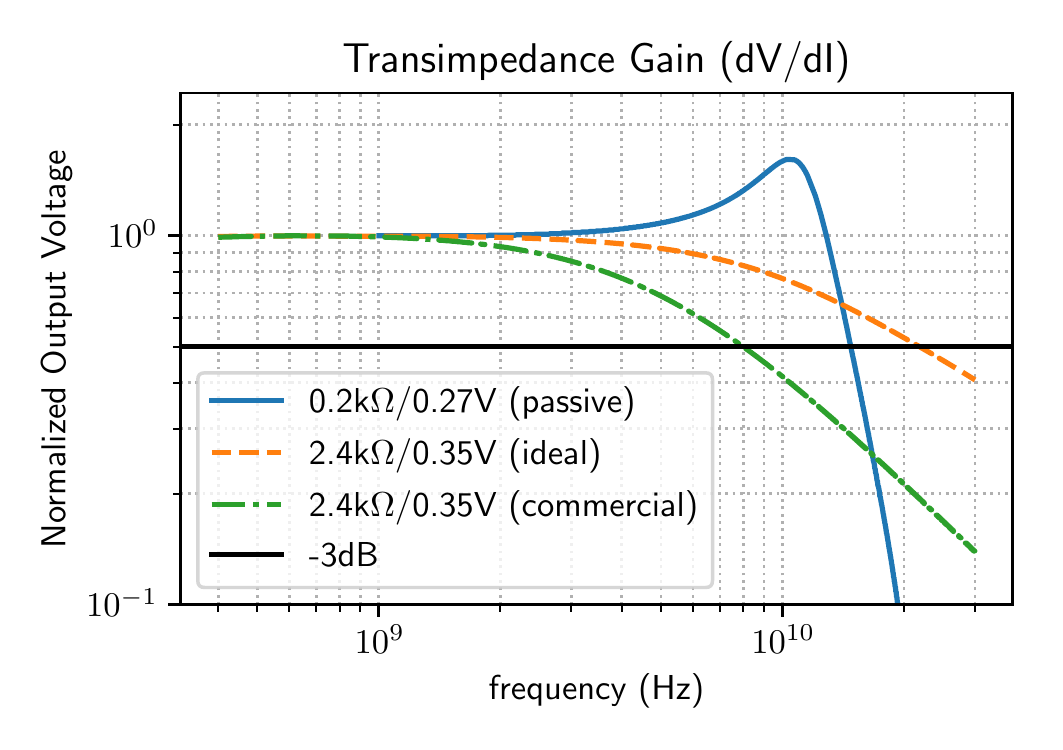}
\caption{Transimpedance gain characteristics of the O/E/O module assuming an AC current source at one of the photodetectors, while measuring AC voltage amplitude across the modulator. The passive transimpedance circuit parameters were introduced in Fig.~\ref{fig:passiveTIA}. The commercial off-the-shelf TIA ONET8531T (Texas Instruments) replaces $R_{\text{TIA}}$ and $L_{\text{peak}}$. The ideal TIA design is similar to that of ref.~\cite{Jun-DeJin2008} but was optimized specifically for this bandwidth and gain range. Note: 
{\color{black} The $f_{\text{3dB}}$ values are: \SI{14.8}{\GHz} (passive), \SI{8.0}{\GHz} (commercial), \SI{21.9}{\GHz} (ideal).} 
The $f_{\text{peak}}$ value for the passive circuit is \SI{11.8}{\GHz}.
}
\label{fig:S21}
\end{figure}

Figure~\ref{fig:S21} shows the bandwidth performance simulation of the O/E/O circuit in Fig.~\ref{fig:passiveTIA} using two active TIA designs (one commercial and one ideal), and how they compare against the passive transimpedance approach. They were all designed to a bandwidth greater than \around\SI{10}{\GHz}. As expected, the use of active TIA allowed us to achieve a \around17 times higher gain-bandwidth product {\color{black} ($\SI{21.9}{\GHz} \times \SI{2.4}{\kilo\ohm}$ vs. $\SI{14.8}{\GHz} \times \SI{0.2}{\kilo\ohm}$)}.

We note that TIAs fabricated with modern CMOS nodes have their maximum output swing voltage limited by the maximum $V_\mathrm{DD}$ of the transistor gates, which, in turn, is limited by the breakdown voltage of the node. For example, the \SI{0.18}{\um}-CMOS node offer TIAs with $2.5\,V_\mathrm{DD}$~\cite{Jun-DeJin2008}, {\color{black} thus limiting the maximum achievable $V_\mathrm{pp}$ to about \SI{1.8}{V}}. Since the modulation depth ($\mathrm{MD}$) is proportional to $V_\mathrm{pp}$, this limit does not impact the gain cascadability condition (Eq.~\ref{eq:gain cascadability}). But {\color{black} if $V_\mathrm{pp} \leq \SI{1.8}{V} < V_\pi$, the modulator will operate in a more linear regime, which will then transmit more noise (higher transmission factor $T_n$),} impacting the noise cascadability condition (Eq.~\ref{eq:SNR}). {\color{black} As a result, neuromorphic photonics would benefit from photodiodes and modulators with driving voltages lower than the $V_\mathrm{DD}$ of modern technology nodes.}

\begin{table}
\centering
\caption{Computed tradeoffs of various O/E/O designs.}
\label{tab:SNRvsP}
\resizebox{\linewidth}{!}{
\begin{tabular}{
r
    c
        c
            c
                c
                    c}
 \toprule
Modulator &
    \multirow{2}{*}{TIA} &
        $V_{\text{pp}} $ &
            $R_{\text{TIA}}$ &
                $P_L$ &
                    SNR (\si{dB}) \\
Class&
    &
        \si{\V} &
            \si{\kilo\ohm} &
                \si{dBm}&
                    $(T_n=0.5)$ \\
\midrule
\multirow{2}{*}{\begin{tabular}{@{}r@{}}p-n junction\\\cite{Xuan2014,Tait:18:mrrMod}\end{tabular}} &
    passive &
        \num{4.8} &
            \num{0.4} &
                \num{26} &
                    \num{64}\\
&
    active &
        \num{4.8} &
            \num{2.8} &
                \num{17} &
                    \num{54}\\
\addlinespace
\multirow{4}{*}{\begin{tabular}{@{}r@{}}graphene \cite{amin2018attojoule}\\\cite[Fig.6]{Goykhman2016,Nozaki2018:nanoPD}\end{tabular}} &
    \multirow{2}{*}{passive} &
        \num{.75} &
            \num{2.0} &
                \num{18} &
                    \num{49}\\
&
    &
        \num{.1} &
            \num{1.2} &
                \num{11} &
                    \num{41}\\
\addlinespace
&
    \multirow{2}{*}{active} &
        \num{.75} &
            \num{2.8} &
                \num{17} &
                    \num{41}\\
&
    &
        \num{.1} &
            \num{2.8} &
                \num{8} &
                    \num{24}\\
 \bottomrule
\end{tabular}
}
\raggedright
~\\
This table was computed by using equations (\ref{eq:gain cascadability}--\ref{eq:SNR}) with the following parameters: $T_n=0.5$; $k_BT = \SI{4.11e-21}{J}$; $\Delta f = \SI{10}{\GHz}$; $q = \SI{1.6e-19}{C}$; $I_{\text{TIA,n}} = \SI{20e-12}{A/\sqrt{Hz}}$ (active), $0$ (passive); $\mathrm{RIN} = \num{1e-6}$~\cite{Resneau2006}; $\mathrm{MD} = 0.61$ (p-n junction), $0.33$ (graphene); $N_{\text{FO}} = 10$; $T_{1/2} = 0.5$; $R_d = \SI{1}{A/W}$ (p-n junction), $\SI{0.35}{A/W}$ (graphene \cite{Goykhman2016}); $\eta_{\text{pp}} = 0.5$; $R_L = R_{\text{TIA}}$ (passive), $R_L = \infty$ (active). $R_{\text{TIA}}$ values were chosen to fulfill $2\pi RC = \Delta f^{-1}$, with capacitances listed in the references.
\vspace{-15pt}
\end{table}


We study the effect of the ideal active TIA in the cascadability condition of the O/E/O neuron.
{\color{black}
Equations~\ref{eq:gain cascadability}~and~\ref{eq:SNR} relate the autapse SNR to multiple design parameters, such as $V_{\text{pp}}$, $P_L$, $R_\mathrm{TIA}$ and bandwidth $\Delta f$. 
There are two main trends to bear in mind. With a passive transimpedance gain, there is a clear tradeoff between bandwidth and power consumption for a fixed modulator design. The larger the bandwidth $\Delta f$, the lower the load $R_\mathrm{TIA}$ has to be, and therefore the larger the optical power $P_L$ to guarantee gain cascadability (eq.~\ref{eq:gain cascadability}). The autapse SNR is constant over values of $\Delta f$, since noise terms are proportional to $R_\mathrm{TIA} \cdot \Delta f$ terms.
On the other hand, an active TIA may provide a higher transimpedance-bandwidth product ($R_\mathrm{TIA} \cdot \Delta f$), but introduces more noise and consumes more power than a passive transimpedance. However, a higher gain allows laser pump power to be lower. Since off-chip lasers tend to be inefficient, this can result in overall system-wide power savings. The TIA also allows for dynamic transimpedance tuning, which is useful to reduce power consumption at times when noise is not critical to the application.

The dependence between SNR and individual parameters shows very simple trends -- linear or quadratic. Therefore, instead of displaying arrays of plots, we chose to use Table~\ref{tab:SNRvsP} to show a few specific examples tied to the devices reported in the literature.
}
The table shows the effect of adding an active TIA to a p-n junction standard-platform modulator vs. a graphene-based ultra-sensitive modulator with $V_\pi$ of \SI{0.1}{V} or \SI{.75}{V}. A standard-platform silicon photonic neuron can benefit much more from the TIA; the laser power requirement decreases by 10-fold {\color{black} (from \SI{26}{dBm} to \SI{17}{dBm}), with the expense of a decrease in the SNR limit (from \SI{64}{dB} to \SI{54}{dB})}. In contrast, almost no enhancement is seen by the graphene-based neuron, because {\color{black} it features low capacitance and therefore} the passive $R_{\text{TIA}}$ {\color{black} can be high without impacting bandwidth}. With that said, TIAs will continue to be useful for applications requiring marrying conventional modulators, and low-power and lower bandwidth budgets, two factors that scale favorably to justify their use.

\section{Conclusion}
\label{sec:3 conclusion}
We presented a quantitative overview of the interplay between electrical gain, optical gain, and modulator sensitivity in O/E/O photonic neurons. Although these metrics have positive influences over the accuracy and stability of the neural network, they can impose difficult power and noise requirements.

We showed that current p-n junction-based silicon photonic platforms do not support highly-interconnected photonic neural networks unless they use (a) more sensitive modulators, (b) active transimpedance amplifiers (TIAs), or (c) operate at a sub-\si{\GHz} bandwidth.

This occurs because modulators need a large voltage swing to reach the nonlinear threshold in their nonlinear transfer function, which suppresses noise directly between one neural layer and the next -- a requirement for cascadable analog links. This swing can either be achieved by increasing optical pump power at the modulator or by providing electric transimpedance gain. However, optical gain is limited by optical nonlinearities in waveguides (and potentially power budgets), whereas transimpedance gain is inversely proportional to the bandwidth of the circuit.

The analysis shown here also applies to other kinds of electro-optic transducers, such as lasers. In particular, excitable lasers~\cite{Prucnal2016} have a zero-or-one thresholding response that serves as both an amplifier and a noise suppression device. Cascadability properties have recently been demonstrated in one such devices~\cite{Peng:19:CLEO}.
Future work will involve testing and quantifying noise cascadability by means of an autapse test experiment, setting another useful benchmark in the field of analog neuromorphic photonics.

\appendices

\newcommand{\avg}[1]{\left\langle#1\right\rangle}
\newcommand{\overbar}[1]{\mkern 1.5mu\overline{\mkern-1.5mu#1\mkern-0.5mu}\mkern 0.5mu}
\renewcommand{\d}[1]{\ensuremath{\operatorname{d}\!{#1}}}

\section{Derivation of Cascadability Equations}
\label{appendix:derivation}

In this section, we will derive the expressions contained in Eqs.~\ref{eq:gain cascadability} and \ref{eq:SNR}.
We will be using the following notations:
\begin{align*}
\overbar{x}(t) \equiv \mathbb{E}\left[x(t)\right];\quad\avg{x(t)} \equiv \int_{0}^{T} x(t) \d t.
\end{align*}
We start by computing the expressions of the photocurrent in Fig.~\ref{fig:noise sources}. The photodetector is often modelled as a linear, added with shot noise and thermal noise terms following normal distributions. This approximation is valid if the optical power amplitude is much slower than the photodiode's response time.

\begin{align}
I = R_d \cdot (P_\text{in} + P_\text{in,n}) + I_\text{shot} + I_\text{thermal}
\end{align}
where
\begin{align}
P_\text{in,n} &\sim \mathcal{N}(0, P_n^2) \\
I_\text{shot} &\sim \mathcal{N}(0, 2q_e \Delta f \overbar{I}) \IEEEnonumber\\
              &\sim \mathcal{N}(0, 2q_e \Delta f R_d \overbar{P_\text{in}})\\
I_\text{thermal} &\sim \mathcal{N}\left(0, \frac{4k_BT}{R_L}\Delta f\right)
\end{align}
are random variables sampled in time. Note: $I_\text{shot}$ should technically not be a normal random variable, but in many textbooks it is modelled that way. It is a good approximation for large values of $\overbar{I} / \Delta f q_e$ (average event count).

Similarly, the transimpedance amplifier is modelled as a linear component with input referred noise.

\begin{align}
V &= R_\text{TIA} \cdot I + V_\text{bias} + V_\text{TIA,n}\\
V_\text{TIA,n} &\sim \mathcal{N}(0, R_\text{TIA}^2 I_\text{TIA,n}^2 \Delta f).
\end{align}
The $I_\text{TIA,n}$ is often presented in the literature in units of \si{\pA/\sqrt{\Hz}} and is assumed invariant for TIAs of the same design on the same technology node. $V_\text{bias}$ is assumed to be noise-free.

So far, it can be easily shown that $V$ is a random variable with Gaussian noise, resulting in

\begin{align}
V &= \overbar{V} + V_\text{noise}\\
\overbar{V} &= R_\text{TIA} R_d  \overbar{P_\text{in}} + V_\text{bias}\\
V_\text{noise} &\sim \mathcal{N}(0, V_n^2).
\end{align}

So far all noise terms have been additive. The next expression for the power envelope of the output optical signal $P_\text{out}$.

\begin{align}
P_\text{out} &= P_L (1 + n_\text{RIN}) \cdot T(V)
\label{eq:appendix:Pout}\\
n_\text{RIN} &\sim \mathcal{N}(0, \text{RIN}^2)
\end{align}
where $T(V)$ is the transfer function of the modulator (Fig.~\ref{fig:mod transfer function}) and RIN is the relative intensity noise -- r.m.s. integrated from 0 to $\Delta f$, $\int|\mathrm{RIN(dBc/Hz)}|\mathrm{d}f$ -- at the pump. There are two nonlinear transformations in this expression that render analytical treatment difficult: RIN is a multiplicative noise term, and $T(\cdot)$ is a nonlinear function. That means that the probability distribution function of the output noise does not have an analytical form. but we can proceed to compute the signal-to-noise ratios at the input and output.

\begin{align}
\text{SNR}_\text{in} &= \frac{\avg{\overbar{P_\text{in}^2}} - \avg{\overbar{P_\text{in}}}^2}{\avg{\overbar{\left(P_\text{in} - \overbar{P_\text{in}}\right)^2}}} = \frac{\avg{\overbar{P_\text{in}^2}} - \avg{\overbar{P_\text{in}}}^2}{P_n^2} \label{eq:appendix:SNRin}\\
\text{SNR}_\text{out} &= \frac{\avg{\overbar{P_\text{out}^2}} - \avg{\overbar{P_\text{out}}}^2}{\avg{\overbar{\left(P_\text{out} - \overbar{P_\text{out}}\right)^2}}}\label{eq:appendix:SNRout}\\
F &= \frac{\text{SNR}_\text{in}}{\text{SNR}_\text{out}} = 1
\label{eq:appendix:F}
\end{align}

The noise cascadability condition is expressed in Eq.~\ref{eq:appendix:F}. The idea is to compute Eq.~\ref{eq:appendix:SNRout} and then express the resulting $P_n^2$ term in terms of $\text{SNR}_\text{in}$ per Eq.~\ref{eq:appendix:SNRin}.

Equation~\ref{eq:appendix:F} is general; it should work for arbitrary amplitude modulation schemes with known probability distribution. From here onwards, we need to choose a convenient input signal waveform that will allow us to compute an approximate expression. We choose an unbiased non-return-to-zero (NRZ) signal as modulation scheme, which is very common in optical links and makes it easy to extract values from experimental papers and plug them into these equations.

\begin{align}
P_\text{in}(t) = P_\text{in,n} +
\begin{cases}
  P_1 & 0 \leq t < T/2 \\
  P_0 & T/2\leq t < T
\end{cases}
\end{align}

We also make a few assumptions about how the modulator is biased, as hinted in Fig.~\ref{fig:mod transfer function}. First, we assume that $V_\text{bias}$ is picked so that
\[
T\left(R_\text{TIA} R_d \cdot \frac{P_1 + P_0}{2} + V_\text{bias}\right) = \frac{1}{2} \equiv T(V_{1/2}).
\]

Second, we denote $T_{0,1} = T(V_{0,1}) = T(V(P_{0,1}))$, $\Delta T \equiv T_1 - T_0$, $\mathrm{MD} \equiv \Delta T / (T_1 + T_0)$, $\Delta P \equiv P_1 - P_0$, and $V_\text{pp} = R_\text{TIA} R_d \Delta P$.

Third, we assume that $T(V)$ is symmetric around $V_{1/2}$ and thus $T'(V_1) = T'(V_0) \equiv T'$ as hinted in Fig.~\ref{fig:mod transfer function}.
It is useful to define the quantity we named the noise transmission $T_n$ as (cf.~Fig.~\ref{fig:mod transfer function}):
\begin{align}
T_n \equiv \frac{T'\cdot V_\text{pp}}{\Delta T}.
\end{align}

Here, we are able to define the gain cascadability condition. We simply state that the output optical amplitude generated by an input amplitude $\Delta P$ should be able to provide at least an amplitude equivalent to $\Delta P$ to each of the fan-out neurons ($N_\text{FO}$). In other words:

\begin{align}
\frac{P_L \Delta T \eta_\text{pp}}{N_\text{FO}} > \Delta P,
\label{eq:appendix:gain cascadability}
\end{align}
where $\eta_\text{pp}$ is efficiency term from point to point, accounting for insertion losses of the WDM networking grid. Eq.~\ref{eq:appendix:gain cascadability} should lead directly to Eq.~\ref{eq:gain cascadability}.

With these assumptions, Eq.~\ref{eq:appendix:SNRin} and the numerator of Eq.~\ref{eq:appendix:SNRout} become, respectively:
\begin{align}
\text{SNR}_\text{in}
&= \frac{\Delta P^2}{4P_n^2} \\
\avg{\overbar{P_\text{out}^2}} - \avg{\overbar{P_\text{out}}}^2
&= \frac{P_L^2\Delta T^2}{4}
\end{align}

In order to complete the derivation, we must know how to approximate the propagation of $P_n^2$ in Eq.~\ref{eq:appendix:Pout}. As shown in Fig.~\ref{fig:mod transfer function}, if the variance of the input signal around their means is small, we can linearize $T(V)$ such that this approximation holds:

\begin{align}
T(V) \approx T(\overbar{V}) + T'(\overbar{V})\cdot V_\text{noise}
\label{eq:appendix:T linear approximation}
\end{align}

Plugging Eq.~\ref{eq:appendix:T linear approximation} into Eq.~\ref{eq:appendix:Pout} and observing that $\overbar{n_\text{RIN}\cdot V_\text{noise}} = 0$ because they are independent, then Eq.~\ref{eq:appendix:SNRout} should become equal to the desired Eq.~\ref{eq:SNR}.
\ifCLASSOPTIONcaptionsoff
  \newpage
\fi



%



\bibliographystyle{IEEEtran}
\bibliography{IEEEabrv,references.bib}

\end{document}